\documentstyle[preprint,aps]{revtex}

\begin{document}
\title{The (1+1)-dimensional Massive sine-Gordon Field
       Theory \\ and the Gaussian Wave-functional Approach}
\author{Wen-Fa Lu }
\address{CCAST(World Laboratory) P.O. Box 8730, Bejing, 100080 \\
         and \\
         Department of Applied Physics, Shanghai Jiao Tong University,
         Shanghai 200030, China
         \thanks{mailing address,E-mail: wenfalu@online.sh.cn}}
\maketitle
\begin{abstract}
The ground, one- and two-particle states of the (1+1)-dimensional massive
sine-Gordon field theory are investigated within the framework of the Gaussian
wave-functional approach. We demonstrate that for a certain region of the
model-parameter space, the vacuum of the field system is asymmetrical.
Furthermore, it is shown that two-particle bound state can exist upon the
asymmetric vacuum for a part of the aforementioned region. Besides, for the
bosonic equivalent to the massive Schwinger model, the masses of the one
boson and two-boson bound states agree with the recent second-order
results of a fermion-mass perturbation calculation when the fermion mass is
small.
\end{abstract}
\vspace{24pt}


\section{Introduction}
\label{1}

The massive sine-Gordon field theory (MsGFT) \cite{1} is a simple
generalization of the massless sine-Gordon field theory (sGFT) \cite{2}, with a
vacuum angle $\theta$ added in the argument of the cosine and a mass term
$m_0^2\phi^2$ added in the Lagrangian. It is well known that the sGFT is
exactly solvable, and can provide a good laboratory for quantum field theory.
Moreover, the (1+1)-dimensional ((1+1)-D) sGFT is equivalent to the massive
$O$(2) non-linear $\sigma$-model, the massive Thirring model, the
two-dimensional Coulomb gas and the continuum limit of the lattice $x$-$y$-$z$
spin-${\frac {1}{2}}$ model. Now this theory has received extensive
investigations \cite{3,4,5}. In the same way, the MsGFT is also an important
model. At any or some special coupling strength, the MsGFT can give a good
description for the dynamics of other important systems, such as the massive
Schwinger model, the Schwinger-Thirring model, the two-dimensional lattice
Abelian Higgs model, the two-dimensional neutral Yukawa gas, and so on
\cite{6,7,8,9}. And again, although it is not yet exactly solved owing to the
existence of the mass term, this model possesses its own field-theoretical
peculiarities \cite{1}, some of which will be discussed in this paper. Hence
it is of general importance to study the MsGFT. Early in 1970's, this theory
was analyzed within the framework of constructive quantum field theory
\cite{1}. So far, as an equivalent system of the massive Schwinger model ( in
this case, the coupling in the MsGFT is only at a special strength
 ), the MsGFT was investigated for large $m_0^2$ by mass perturbation or some
light-cone quantization methods \cite{6,10,11,12}. In order to reveal the phase
structure of the Abelian Higgs model, the MsGFT with a finite momentum cutoff
was treated by renormalization-group technique \cite{8} (1994). Obviously, a
further investigation of the MsGFT (especially at any finite value of the
coupling) is still necessary and of universal usefulness.

In this paper, using the Gaussian wave-functional approach (GWFA), we intend to
investigate the (1+1)-D MsGFT with zero vacuum angle $\theta=0$ at any coupling
strength. The Lagrangian is
\begin{equation}
{\cal L}={\frac {1}{2}}\partial_\mu \phi_x \partial^\mu \phi_x-
                      {\frac {1}{2}}m_0^2\phi_x^2-
          {\frac {m^2}{\beta^2}}(1-cos(\beta\phi_x)) \equiv
          {\frac {1}{2}}\partial_\mu \phi_x \partial^\mu \phi_x-U(\phi_x),
\end{equation}
with $\phi_x\equiv\phi(x)$, where $m_0$ and $m$ are in mass dimension and the
dimensionless $\beta$ is the coupling parameter. It is always viable to have
$\beta^2\ge 0$ \cite{2}. In the case of $m_0=0$, Eq.(1) describes the sGFT, and
when $\beta^2\to 0$, Eq.(1) describes a free theory of the squared mass
($m_0^2+m^2$)(if $m_0^2+m^2>0$). Evidently, the Lagrangian Eq.(1) is invariant
under the transformation of $\phi\to (-\phi)$ . We shall be particularly
interested in the spontaneous symmetry breakdown (SSB) and the two-particle
bound states upon the SSB vacuum ( In this paper, by SSB, we mean that the
energy at a symmetric vacuum with $\phi=0$ is exactly higher than at an
asymmetric vacuum with $\phi\ne 0$ ) \footnote{Generally, symmetry breakdown
also includes the phenomenon that the energy at a symmetric vacuum with $\phi=0$
is exactly equal to the one at an asymmetric vacuum with $\phi\ne 0$. This
phenomenon is called degeneration in this paper.}. Also, we shall compare our
results about the masses of the one particle and the two-particle bound states
with the ones in the literature.

We hope to demonstrate qualitatively the existence of the SSB in the MsGFT. As
is known, the classical potential of the sGFT is invariant under the
transformation of $\phi\to (\phi+{\frac {2n\pi}{\beta}})$ ( hereafter $n$ is an
integer). Hence the classical vacua of the sGFT are infinitely degenerate, and
the corresponding quantized vacua are degenerate likewise \cite{3,13}. But for
the MsGFT, the situation is quite different. A simple analysis indicates that
because of the existence of the mass term, the classcal vacuum of the MsGFT is
unique at $\phi=0$ for a negative $m^2$ with $|m^2|<m_0^2$ or for a positive
$m^2$, whereas it is located at $\phi\neq 0$ for a negative $m^2$ with
$|m^2|>m_0^2$ ( In the case of $\beta^2=4\pi$, this is compatible with
Ref.~\cite{12}. Note that in the caption of fig.5 of Ref.~\cite{12}, the word
``large'' should perhaps be the words with the meaning ``sufficiently
negative'', according to the context there ). As suggested in Ref.~\cite{12},
the SSB is usually believed to be kept by the corresponding quantized vacuum.
Undoubtedly, this phenomenon is interesting and useful both for electroweak
theory and for the above equivalent models. Nevertheless, the investigations of
the MsGFT in the literature were mostly achieved for a small $m^2$ and made no
explicit investigations of the quantum SSB ( to our knowledge ). In the next
section, we shall demonstrate the existence of the SSB and give its relevant
region in the model-parameter space.

There exist a large variety of bound state systems in nature, but investigation
of them is a hard task in quantum field theory \cite{14}. For this task, the
GWFA is an effective and feasible tool in practice and can give qualitatively
correct results about bound states. Up to now, within the framework of the
GWFA, bound state has been shown to exist upon symmetric vacuum of the
following models : the (1+1)-D $:\lambda(\phi^6-\phi^4):$ model \cite{15}, the
$\phi^6$ theory in $(1+1)$ and ($2+1$) dimensions \cite{16}, the Gross-Neveu
model \cite{17}, the sGFT and the double sine-Gordon model \cite{13,18}.
However, the GWFA has not established bound states upon the SSB vacuum of any
model as yet. Although the $\lambda\phi^6$ theory has typically the SSB
phenomenon, the interaction between the two particles is repulsive in this case
and therefore the two particles form impossibly a bound state upon the SSB
vacuum \cite{16}. Perhaps the MsGFT can give an example of such a phenomenon.
Section III will concentrate on it and one will see that the story is really
so.

In the last decade, the GWFA \cite{19} has become a powerful tool to extract
the non-perturbative information of many field theoretical models
\cite{15,16,17,20}. To be true, there have two unfavourable facts for the GWFA.
One is that the GWFA gives the wrong order of the phase transition in the
$\lambda \phi^4$ theory \cite{21,23}, and another is that it is difficult to
control the approximate accuracy of the GWFA. Nevertheless, endeavours in the
last decade have led to a little progress in controlling the accuracy.
\cite{22}. Moreover, the GWFA predicts correctly the existence of the phase
transition \cite{23,24} which maybe have second-order feature \cite{23,22},
albeit it wrongly predicts the first order of transition for some (1+1)-D
quantum field theories. A great deal of the existing work has shown that the
GWFA is a tractable and helpful non-perturbative tool. It can give a
qualitatively correct information \cite{21} or a precursory study at least
\cite{25,22}. In the present work, we hope to give a further helpful support
to the GWFA through comparing our results with those obtained from the massive
Schwinger model. As mentioned in the above, the system Eq.(1) at a special
coupling strength ( $\beta^2=4\pi$ ) is equivalent to the massive Schwinger
model at zero charge sector, in which there are only Schwinger boson
( fermion-antifermion bound state) and its bound states \cite{6}(Coleman).
Hence, the MsG bosons at $\beta^2=4\pi$ are just Schwinger bosons. Last two
decades, the masses of Schwinger boson and its bound states have been
calculated at a good accuracy with the aid of many methods
\cite{11,26,27,28,29,30}, most of which were numerically or analytically
completed in the limit of the strong and/or weak coupling but not at all values
of the coupling. On the other hand, our results about the MsGFT are analytical
at any set of values of the model parameters. Hence a comparison between ours
and the others may be made. This will be done mainly for small fermion mass in
Section IV. One will see that when $\beta^2=4\pi$, for the large $m_0^2$,
$i.e.$, for small fermion mass parameter, the GWFA results for the masses of
the Schwinger boson and its bound state have a good agreement with those in
Refs.~\cite{26,27,28,29,30}.

In the next section, we shall directly give the Gaussian effective potential
( GEP ) of the MsGFT and then discuss vacuum structure. As for the
procedure of the GWFA, there have been detailed discussions in many references,
such as Refs.~\cite{4,19,20,21,31}, and we intend to have no more explanations
in this paper. Section III is devoting to the excited states. For both the
symmetric and the asymmetric vacua in section II, we shall analyze existence of
two-MsG-boson bound states and calculate the masses of the one MsG boson and
two-MsG-boson bound state. In section IV, the masses of the one Schwinger boson
and two-Schwinger-boson bound state in the massive Schwinger model will be
given from the results of the MsGFT by employing the equivalence between the
two models,, and compared with those in recent Refs.~\cite{29,30}. A brief
conclusion and some discussions will be made at the end of this paper.

\section{Vacuum Structure and Stability}
\label{2}

This section consider the ground state in the system Eq.(1).

In Eq.(1), we have to maintain a positive $m_0^2$ for avoiding an
unbounded-below vacuum. Nevertheless, different from the sGFT,
both the positive and the negative $m^2$ should be considered in Eq.(1) because
the physics of the negative is not equivalent to the one of the positive.
Moreover, as stated in the last section, the classical vacuum of the system
Eq.(1) is infinitly degenerate no longer. It is symmetrical for a negative
$m^2$ with $|m^2|<m_0^2$ or positive $m^2$, and become asymmetrical when $m^2$
is negative enough. In this section, we intend to investigate the structure and
properties of the quantum vacuum through the GEP.

In the fixed-time functional Schr\"odinger picture, the normal-ordered
Hamiltonian operator corresponding to Eq.(1) is
\begin{eqnarray}
&{\cal N}_M[H]=&\int_x \bigl[ {\frac {1}{2}}\Pi^{2}_x
                +{\frac {1}{2}}(\partial_x \phi_x)^2
                +{\frac {1}{2}}m_0^2\phi_x^2 -{\frac {1}{4}}m_0^2I_1(M^2)
                -{\frac {1}{2}}I_0(M^2) \nonumber \\ &\ \ \ &
                + {\frac {1}{4}}M^2I_1(M^2)
                -{\frac {m^2}{\beta^2}}{\cal N}_M[cos(\beta\phi_x)]
                exp\{-{\frac {\beta^2}{4}}I_1(M^2)\} +
                {\frac {m^2}{\beta^2}} \bigr]  \;,
\end{eqnarray}
with the notation
\begin{center}
$I_n(M^2)
  =\int {\frac {dp}{2\pi}}{\frac {\sqrt{p^2 + M^2}}
{(p^2 + M^2)^n}} \;$  .
\end{center}
Here, $\Pi_x\equiv -i{\frac {\delta}{\delta \phi_x}}$ is conjugate to the field
operator $\phi_x$, $\int_x\equiv\int dx$ the integration in 1-dimensional
coordinate space, and ${\cal N}_M[\cdots]$ means normal-ordering with respect
to any positive constant $M$ ($M$ is with mass dimension and usually called
normal-ordering mass). Take as an ansatz the general Gaussian wave functional
\cite{4,19,31}
\begin{equation}
|\Psi>\to \Psi[\phi;\varphi,{\cal P},f] = N_f exp\{i\int_x{\cal P}_x\phi_x
      -{\frac {1}{2}}\int_{x,y}(\phi_x - \varphi_x)f_{xy}(\phi_y - \varphi_y)\}
\end{equation}
with $N_f$ some normalization factor, and ${\cal P}_x$, $\varphi_x$ as well as 
$f_{xy}$ being the variational parameter functions. Using functional
integration techniques \cite{32,4,33}, one can first calculate the energy
$\int_x<\Psi|{\cal N}[{\cal H}_x]|\Psi>$, then take $\varphi_x$ as a constant
$\varphi$, and finally minimize variationally the energy in respect to
${\cal P}$ as well as $f$. Consequently, ${\cal P}_x=0$, the Fourier component
of $f_{xy}$ is
$$f(p)=\sqrt{p^2 + \mu^2(\varphi)}$$
and the GEP reads
\begin{eqnarray}
V(\varphi)&=&{\frac {1}{2}}[I_0(\mu^2)-I_0(M^2)]
            -{\frac {1}{4}}[\mu^2I_1(\mu^2)-M^2I_1(M^2)]  \nonumber \\
            &\;& +{\frac {1}{4}}m_0^2[I_1(\mu^2)-I_1(M^2)]
            +{\frac {1}{2}}m_0^2\varphi^2     \nonumber \\
            &\;&  +{\frac {m^2}{\beta^2}}[1-exp\{-{\frac {\beta^2}{4}}
            [I_{1}(\mu^2)-I_1(M^2)]\}cos(\beta\varphi)] \;.
\end{eqnarray}
Here, $\mu$ takes one of the following three possible values : the nonzero
root of the gap equation
\begin{equation}
\mu^2=\mu^2(\varphi)= m_0^2+m^2 exp\{-{\frac {\beta^2}{4}}
         [I_{1}(\mu^2)-I_1(M^2)]\}cos(\beta\varphi) \;,
\end{equation}
the two end points of the range $0\le\mu^2<\infty$ $\mu^2=0$ and $\mu^2\to
\infty$ (the explanation of this point is put off to the next paragraph). In
the $r.h.s.$ of Eq.(5), $\mu$ is a function of the uniform background field
$\varphi$, which is the vacuum expectation of the field operator $\phi$. Among
the minimized results, ${\cal P}_x$ is the average value of the total momentum
density operator of the field system, and its null result is understandable. As
for $f(p)$ and $\mu$, its physical meaning will get transparent in the next
section. (By the way, because $U(\phi)$ has its Fourier representation in a
sense of the tempered distribution \cite{34}, the above GEP and the energies of
the one- and two-particle states in the next section can be also calculated
directly as per formulae in Ref.~\cite{31}, for which it is enough to finish
some simple integrals).

Noticing the following results of the integrals
  $${\frac {1}{2}}[I_1(\mu^2)-I_1(M^2)]
  =-{\frac {1}{4\pi}}ln{\frac {\mu^2}{M^2}}$$
  and
$${\frac {1}{2}}[I_0(\mu^2)-I_0(M^2)]
  -{\frac {1}{4}}[\mu^2I_1(\mu^2)-M^2I_1(M^2)]={\frac {\mu^2-M^2}{8\pi}} \;,$$
one can see that Eqs.(4) and (5) contain no divergences, and therefore, further
renormalization procedure is not necessary. Nevertheless, in order to compute
the GEP from Eq.(4), we have to choose the value of $\mu$ among the three
possible values : 0, $\infty$ and the nonzero-root of Eq.(5). The existence of
the three possible values is because $\mu(\varphi)$ in Eq.(5) results from the
process of minimizing the energy with respect to $f$ ( $\mu(\varphi)$ in Eq.(5)
is the stationary point ) and the GEP must be the global minimum of the energy
density for the whole range of $\mu^2$ ( $0\le\mu^2<\infty$ ) \cite{16,21,24}.
Therefore, for every value of $\varphi$, one has to compare
${\cal V}(\varphi)$'s at the three possible values of $\mu$ with each other,
and only the minimum among them can be taken as the GEP.

Obviously, the end point $\mu=0$ enforces $V(\varphi)$ in Eq.(4) infinite and
must be discarded. Moreover, for the other end point $\mu\to\infty$, one has
$V(\varphi)\to {\frac {\mu^2}{8\pi}}-{\frac {m^2}{\beta^2}}
    ({\frac {\mu^2}{M^2}})^{\frac {\beta^2}{8\pi}}cos(\beta\varphi) . $
This implies that when $\beta^2<8\pi$, $V(\varphi)\to {\frac {\mu^2}{8\pi}}$
and tends to infinite for the infinite $\mu$. Thus when $\beta^2<8\pi$, one
should resort to only the nonzero solution of Eq.(5) for governing the GEP,
which renders $V(\varphi)$ finite. As for the case of $\beta^2>8\pi$, we have
$V(\varphi)\to -{\frac {m^2}{\beta^2}}
       ({\frac {\mu^2}{M^2}})^{\frac {\beta^2}{8\pi}}cos(\beta\varphi).$
For those values of $\varphi$ with $m^2 cos(\beta\varphi)>0$, the end point
$\mu\to \infty$ makes $V(\varphi)$ unbounded from below, and accordingly the
vacuum is unstable. So $\beta^2$ should be smaller than $8\pi$. This constraint
of $\beta^2$ is consistent with that in Ref.~\cite{1} ( the 7th paragraph on
page 372 in the book ), and is a little similar to that in the sGFT \cite{2,4}
( the possible difference about the physical sense of the constraint will be
discussed in section V ). In a word, for computing the GEP, we
should use the nonzero root of Eq.(5) instead of the values $\mu^2=0$ and
$\mu^2\to \infty$, and meanwhile, the coupling parameter $\beta^2$ is
constrained to the range of $0\le\beta^2<8\pi$ ( For the sG systems in
condensed matter physics, the constraint can be extended to $\beta^2<16\pi$
\cite{35}. Perhaps, the constraint $\beta^2<8\pi$ could have some analogous
extension for the MsG systems in condensed matter physics. \footnote{To
determine this point needs some other investigation.} ).

Furthermore, in order to analyze vacuum structure and statbility, we still need
the extremum condition ( ${\frac {d V(\varphi)}{d\varphi}}=0$ )
\begin{equation}
m_0^2\varphi+ {\frac {m^2}{\beta}}exp\{-{\frac {\beta^2}{4}}
         [I_{1}(\mu^2)-I_1(M^2)]\}sin(\beta\varphi)=0
\end{equation}
and the stability condition ( $i.e.$, the second derivative of
$\int_x<\Psi|{\cal N}[{\cal H}_x]|\Psi>$ with respect to the relevant
variational parameter $f$ must be positive \cite{4,21,24} at $\mu(\varphi)$ )
\begin{equation}
1-{\frac {m^2\beta^2}{8}}exp\{-{\frac {\beta^2}{4}}
         [I_{1}(\mu^2)-I_1(M^2)]\}I_2(\mu^2)cos(\beta\varphi))>0 \;.
\end{equation}

When $\beta^2<8\pi$, Eq.(5) always has a nonzero solution ( which is different
from the sGFT, where no solutions can exist for some values of $\beta\varphi$
\cite{4} ), and accordingly we can define a parameter with mass dimension
\begin{equation}
\mu_0^2\equiv \mu^2(\varphi=0)= m_0^2
       +m^2 exp\{-{\frac {\beta^2}{4}}[I_{1}(\mu_0^2)-I_1(M^2)]\} \;,
\end{equation}
which is positive ( independent of the sign of $m^2$ ), and is physical mass
squared when vacuum is symmetric ( see the next section ). When $m^2<0$,
$\mu_0<m_0$, otherwise $\mu_0>m_0$. In Eqs.(4),(5),(6) and inequality (7), we
can eliminate $m^2$ in favor of $\mu_0^2$ with the help of the definition (8).
For the convenience of numerical computation, we define the following
dimensionless quantities
\begin{equation}
{\tilde{V}}(\varphi)\equiv {\frac {V(\varphi)-V(\varphi=0)}{\mu_0^2}},
{\tilde{\mu}}\equiv {\frac {\mu}{\mu_0}},
{\tilde{m}}_0\equiv {\frac {m_0}{\mu_0}} \;.
\end{equation}
Then, we can rewrite the GEP as
\begin{equation}
\tilde{V}(\varphi)
=({\frac {1}{8\pi}}-{\frac {1}{\beta^2}})({\tilde{\mu}}^2-1)-
  {\frac {\tilde{m}_0^2}{8\pi}}ln(\tilde{\mu}^2)
  +{\frac {1}{2}}\tilde{m}_0^2\varphi^2    \;,
\end{equation}
the gap
\begin{equation}
\tilde{\mu}^2=\tilde{m}_0^2+(1-\tilde{m}_0^2)(\tilde{\mu}^2)^
      {{\frac {\beta^2}{8\pi}}}cos(\beta\varphi)) \;,
\end{equation}
the extremum condition
\begin{equation}
\tilde{m}_0^2\varphi+{\frac {\tilde{\mu}^2-\tilde{m}_0^2}{\beta}}
         tan(\beta\varphi)=0
\end{equation}
and the stability condition
\begin{equation}
1-\beta^2{\frac {\tilde{\mu}^2-\tilde{m}_0^2}{8\pi\tilde{\mu}^2}}
         >0 \;.
\end{equation}
Here, a divergent constant has been discarded in Eq.(10). Thus, the calculation
of the GEP from the last four equations or inequality is equivalently carried
out at a fixed value of $m^2$. When one intends to consider the effects of
$m^2$, it is enough to further utilize the original definition Eqs.(8) and (9).
( Of course, in order to define dimensionless quantities, one can have other
choices. For example, one can choose the parameter $m_0^2$ as a unit instead of
$\mu_0$. )

Now we can compute the GEP of the MsGFT. It is difficult to solve analytically
Eqs.(10)---(12) and the inequality (13), and hence we have to appeal to a
numerical method for tackling them. If there exists an SSB phenomenon, the
analysis of the vacuum structure indeed amounts to the determination of the
boundary between the symmetric and asymmetric vacua in the model-parameter
space $\beta^2$---$\tilde{m}_0^2$. In order to obtain the boundary, one can
search for the points in the space $\beta^2$---$\tilde{m}_0^2$ at each of which
the value of GEP for the asymmetric vacuum ($\varphi\ne 0$) is exactly equal to
the one for the symmetric vacuum. When one excutes the numerical computation,
an additional point to be noticed is that for some values of the parameters
\{$\beta^2, \tilde{m}_0^2$\} there exist two roots of Eq.(11) and one of them
should be chosen so as to minimize $\tilde{V}(\varphi)$ in Eq.(10).

The numerical results indicate that: ($i$). when $\tilde{m}_0^2<2$ ( an
approximate value ), the vacuum is symmetrical; ($ii$). for any
$\tilde{m}_0^2>2$, there is a critical value of the coupling parameter
$\beta_c^2$, at which the vacuum is degenerate and located at either
$\varphi=0$ or $\varphi\not=0$. When $\beta^2<\beta_c^2$ the vacuum is
symmetrical, whereas when $\beta^2>\beta_c^2$, the symmetry of the vacuum is
broken. Collecting all the above, we depict the $\tilde{m}_0^2$---$\beta^2$
parameter space as Fig.1. The allowed region of the parameters at any fixed
$\mu_0$ forms a semi-infinite strip \{$\tilde{m}_0^2>0, 0\le\beta^2<8\pi$\},
and in this strip, the long-dashed curve represents the critical coupling
$\beta_c^2$ ( The dotted  and short-dashed curve are relevant to bound states
and will be explained in the next section.). In Fig.1, the domain $I$
corresponds to the symmetric vacuum, and the domains $II$ and $III$ correspond
to the asymmetric vacuum. From this figure, one can see that with the increase
of ${\tilde{m}_0}$ the domain $I$ gets more and more narrow. That is to say,
the more negative $m^2$ is, the wider the domain of $\beta^2$ for the
asymmetric vacuum is. For an illustrative purpose, we plot the GEP in Figs.2
and 3 for $\tilde{m}_0^2=1.5$ and $\tilde{m}_0^2=20$, respectively. For the
latter, $\beta_c^2 \approx 3.0265772$ ( Note that the GWFA value of $\beta_c^2$
could not predict an exact value of the critical point when a phase transition
near $\beta_c^2$ is considered, and perhaps the value of $\beta^2_c$ and some
relevant information could be changed by some better approximate approach. The
discussion relevant to this point will be defered to section V. ). Thus, we see
that when $\tilde{m}_0^2>2$, there really exists an asymmetric vacuum within
the framework of the GWFA. That is to say, for the MsGFT, the classical
double-well potential can causes an SSB in the quantum theory. This is usually
believed, just as pointed out in Ref.~\cite{12}. When
$\beta^2<{\frac {16}{\pi}}$, this is also compatible with Ref.~\cite{1} ( page
382 in the book ) .

  About the asymmetrical vacuum of the MsGFT, we have more story to mention.
Ref.~\cite{1} ( the Letter ) pointed out that for $m^2$ large enough and
positive, the $\phi\to (-\phi)$ symmetry is presumably dynamically broken. This
implies that for a sufficiently small $\tilde{m}_0^2$, the vacuum can be
asymmetrical. Nevertheless, using the above GWFA results, we failed to find a
very small but nonzero value of $\tilde{m}_0^2$ ( with $\beta^2<8\pi$ ) which
can lead to a dynamic symmetry breakdown ( In order to consider it, we also
chose $m_0^2$ as a unit to perform the numerical calculation and failed
likewise.). Of course, when $\tilde{m}_0^2$ tends to zero, the GWFA can well
give the effective potential for $\beta\varphi\in [(2n-{\frac {1}{2}})\pi,
(2n+{\frac {1}{2}})\pi]$ ( in the case of $m^2>0$ ) \cite{4}, and the vacua are
degenerate, which is a special case of the dynamical symmetry breakdown. Thus,
in Fig.1, the left, linear boundary of the domain $I$ ( $\tilde{m}_0^2=0,
0\ge \beta^2<8\pi$ ) corresponds to the special dynamical-symetry-breakdown
vacuum, which can be either symmetrical or asymmetrical at each point of the
boundary. \footnote{We thank the referees for their comments leading to our
notice about this point.}

Additionally, we want to mention the symmetry restoration by quantum effects
\cite{3,36,24}. From Eqs.(8) and (9), $\tilde{m}_0^2>1$ means $m^2<0$, and thus
the above symmetry-vacuum domain with $\tilde{m}_0^2>1$ in Fig.1 shows the
occurrence of symmetry restoration phenomenon, because the corresponding
classical vacuum can be spontaneously symmetry-broken.

In this section, we have obtained vacuum structure of the MsGFT. The
ground state wave-functional is the best trial Gaussian functional
$\Psi[\phi;\varphi,{\cal P},f]$ in Eq.(3) with ${\cal P}_x=0$,
$\varphi_{x}=\varphi_0$, and the Fourier component of $f_{xy}$ is
$f(p)=\sqrt{p^2 + \mu^2(\varphi_0)}$. Here $\varphi_0$ is a constant satisfied
Eq.(6), at which the GEP is lowest and the vacuum is located. In addition,
there is a constraint of $\beta^2$, that is, $0\le\beta^2<8\pi$. In the next
section, we shall discuss the excited states upon the ground state. For
convenience, we use $|\varphi_0>$ to represent the ground state wave-functional
hereafter.

\section{Bound States}
\label{3}

In this section, we investigate the one- and two-particle excited states.
Following Refs.~\cite{15,16,19,31,33}, one can manufacture the annihilation and
creation operators with respect to the vacuum state $|\varphi_0>$
\begin{equation}
A_f(p) = ({\frac {1}{4\pi f(p)}})^{1/2}\int_x
  e^{-ipx}[f(p)(\phi_x - \varphi_0)
+{\frac {\delta}{\delta\phi_x}}]
\end{equation}
and
\begin{equation}
A^\dagger_f(p) = ({\frac {1}{4\pi f(p)}})^{1/2}\int_x e^{ipx}
[f(p)(\phi_x - \varphi_0) - {\frac {\delta}{\delta\phi_x}}]  \;.
\end{equation}
It is evident that $A_f(p)|\varphi_0>=0$ and the commutator
$[A_f(p),A^\dagger_f(p')]=\delta(p-p')$. Then one has the one-particle state
\begin{equation}
|1> = A^\dagger_f (p)|\varphi_0>
\end{equation}
and the S-wave two-particle state
\begin{equation}
|2>
= \int dp\Sigma(p)A^{\dagger}_f(p)
A^{\dagger}_f (-p)|\varphi_0> \;,
\end{equation}
where $\Sigma(p)$ is the Fourier transformation of the S-wave function of the
two-particle system.

For the one-particle state, one can find
\begin{equation}
m_1={\frac {<1|{\cal N}_M[H]|1>}{<1|1>}}-\int_x {\cal V}(\varphi_0)
=\sqrt{p^2+\mu^2(\varphi_0)} \;,
\end{equation}
which is the energy of one particle with a momentum $p$. This is the physical
sense of $f(p)$, which has previously appeared in the last section. Obviously,
$\mu(\varphi_0)$ is $m_R$ the physical mass of a particle according to the
relevant vacuum. Thus, within the framework of the GWFA, the single-particle
mass of the MsGFT is
\begin{equation}
m_R^2= m_0^2+m^2({\frac {m_R^2}{M^2}})^{\frac {\beta^2}{8\pi}}
        cos(\beta\varphi_0) \;.
\end{equation}
A further analysis tells us that for both the asymmetric and the symmetric
vacua, $m_R$ increases with the increase of $m_0^2$ or $m^2$; for the symmetric
vacuum, $m_R$ also increases when $\beta^2$ increases, but for the asymmetric
vacuum, the story is a little complicated, which here we intend to discuss no
longer.

Now we turn to discuss the two-particle state. From Ref.~\cite{33}, the
two-particle energy $m_2={{<2|H|2>}\over {<2|2>}}-\int_x {\cal V}(\varphi_0)$
can be calculated as ( expressed in terms of the dimensionless quantities )
\begin{equation}
\tilde{m}_2 = {\frac {2\int d\tilde{p} (\Sigma(\tilde{p}))^2 f(\tilde{p})
- {\frac {\beta^2}{16\pi}}(\tilde{\mu}^2(\varphi_0)-\tilde{m}_0^2)
(\int d\tilde{p}{\frac {\Sigma(\tilde{p})}{f(\tilde{p})}})^2}
{\int d\tilde{p}(\Sigma(\tilde{p}))^2}} \;,
\end{equation}
with $\tilde{m}_2\equiv {\frac {m_2}{\mu_0}}$, $\tilde{p}\equiv
{\frac {p}{\mu_0}}$, and $f(\tilde{p})\equiv \sqrt{\tilde{p}^2+
\tilde{\mu}^2(\varphi_0)}$. The two terms in this expression can be regarded as
the kinetic energy of the two constituent particles and their interacting
energy, respectively. Obviously, the interacting energy is closely related to
$(\tilde{m}_0^2-\tilde{\mu}^2 (\varphi_0))$. When $\tilde{\mu}^2 (\varphi_0)
<\tilde{m}_0^2$, the interacting energy is positive, the two particles repel
each other and can not combine into a bound state, while for $\tilde{\mu}^2
(\varphi_0)>\tilde{m}_0^2$, the interacting energy is negative, the two
particles attract each other and may form a bound state.

Analyzing Eq.(11), one can find that for symmetry vacuum ( $\varphi_0=0$ ),
$i.e.$, for the domain $I$ in Fig.1, when $\tilde{m}_0<1$
$\tilde{\mu}^2 (\varphi_0)>\tilde{m}_0^2$ and when $\tilde{m}_0>1$
$\tilde{\mu}^2 (\varphi_0)<\tilde{m}_0^2$; for a symmetry-broken vacuum
( $\varphi_0\not=0$ ), $i.e.$, for the domains $II$ and $III$ in Fig.1, if
$cos(\beta\varphi_0)>0$ then $\tilde{\mu}^2 (\varphi_0)<\tilde{m}_0^2$, and if
$cos(\beta\varphi_0)<0$ then $\tilde{\mu}^2 (\varphi_0)>\tilde{m}_0^2$.
It is worth while noticing that the case of $cos(\beta\varphi_0)<0$ does exist
when $\tilde{m}_0^2$ is less than about 9.5 or when $\beta^2$ is greater than
some value $\beta_b^2$ for any larger $\tilde{m}_0^2$, which is involved in the
domain $III$ in Fig.1. In Fig.1, the short-dashed curve corresponds to the
critical case of $cos(\beta_b\varphi_0)=0$. Thus, for the symmetry vacuum with
$\tilde{m}_0>1$ or for the asymmetry vacuum with $cos(\beta\varphi_0)>0$ (the
domain $II$) the two-particle states can be just the scattering ones, whereas
for the symmetry vacuum with $\tilde{m}_0<1$ or for the asymmetry vacuum with
$cos(\beta\varphi_0)<0$ (the domain $III$), there can exist the two-particle
bound states.

The mass of the bound state $m_b$ can be calculated within the framework of the
GWFA \cite{15,16,13}. Minimizing the energy $\tilde{m}_2$ with respect to
$\Sigma(\tilde{p})$ yields the equation for $\tilde{m}_2$
\begin{equation}
{\frac {\beta^2}{16\pi}}(\tilde{\mu}^2(\varphi_0)-\tilde{m}_0^2)
\int{d\tilde{p}\over f^{2}(\tilde{p})(2f(\tilde{p})-\tilde{m}_2)}=1 \;.
\end{equation}
When $\tilde{\mu}^2(\varphi_0)<\tilde{m}_0^2$ this equation has no solution
but one can obtain the scattering phase shifts \cite{33} \cite{5}(1993). When
$\tilde{\mu}^2(\varphi_0)>\tilde{m}_0^2$, Eq.(21) has a solution with
$\tilde{m}_2<2\tilde{\mu}(\varphi_0)$, and $\tilde{m}_2$ in this equation times
$\mu_0$ is just the mass of the bound state $m_b$. Finishing the integration in
Eq.(21) leads to
\begin{equation}
\tilde{m}_b={\frac {\beta^2}{8\pi}}
    (1-{\frac {\tilde{m}_0^2}{\tilde{\mu}^2(\varphi_0)}})
   [{\frac {1}{\sqrt{1-\tilde{m}_b^2}}}
   tan^{-1}\sqrt{{\frac {1+\tilde{m}_b^2}{1-\tilde{m}_b^2}}}
   -{\frac {\pi}{4}}]
\end{equation}
with the reduced mass $\tilde{m}_b\equiv {\frac {m_b}{2\mu(\varphi_0)}}$.
When $\tilde{m}^2_0<1$, the vacuum is symmetric, $\tilde{\mu}(\varphi_0)$ is
unit, and the last equation is enough to give the reduced mass of the bound
state in the symmetric vacuum. For some values of $\tilde{m}^2_0$, we give the
dependence of the reduced mass upon the coupling constant $\beta^2$ in Fig.4.
This figure indicates that the reduced mass of bound state decreases with the
increase of $\beta^2$, and increases with the increase of $\tilde{m}_0^2$.
( Note that for the symmetry vacuum case, Eq.(22) can give vanishing or even
negative $\tilde{m}_b$ if the curves in Fig.4 is not artificially cut off at
$\beta^2=8\pi$.) For a given asymmetric vacuum, one can calculate the reduced
mass of the bound state through Eqs.(10)---(13) and (22). In this case, a
vacuum located at a different $\varphi_0$ corresponds to a different curve in
the domain $III$ of Fig.1 (Of course, so does in the domain $II$). For
instance, the vacuum at $\varphi_0=0.75$ corresponds to the dotted curve in the
domain $III$. Vacua at other $\varphi_0$ correspond to other similar curves.
For $\varphi_0=0.75$, we depict the dependence of the reduced mass of bound
state upon $\beta^2$ in Figs.5 and upon $\tilde{m}_0^2$ in Fig.6. From Fig.5,
the reduced mass of bound state decreases with the increase of $\beta^2$, which
is similar to that for symmetric vacuum, but is almost fixed at some
not-too-small value ( 0.76 or so ) before $\beta^2$ arrives at the limit
$8\pi$. The dependence upon $\tilde{m}_0^2$ is slightly complex from Fig.6.
When two particles are not tightly bound, the reduced mass of the bound state
increases with the increase of $\tilde{m}_0^2$, and approaches the value 1 (two
free-particle case) when $\tilde{m}_0^2$ rises such a value that the dotted
curve in Fig.1 falls down on the short-dashed curve, which means that
$\tilde{m}_0^2$ acts as the coefficient of the free term in the Lagrangian. On
the other hand, if the binding between two particles is slightly tighter, the
reduced mass decreases with the increase of $\tilde{m}_0^2$, which means that
$\tilde{m}_0^2$ acts as the coefficient of the cosine interaction term in the
Lagrangian, and keeps at the fixed value 0.76 or so for larger $\tilde{m}_0^2$.
The monotonous decrease of $\tilde{m}_b$ upon $\tilde{m}_0^2$ is understandable
because $(1-\tilde{m}_0^2)$ plays the role of ${\frac {m^2}{\mu^2_0}}$ in the
reduced expression Eq.(11). These results indicate that from Eq.(22) the bound
state in an asymmetric vacuum becomes never ultratightly bound.

In this section, we have obtained the single-particle mass of the MsGFT, and
shown that for both the symmetric and the asymmetric vacua, there exist
two-particle bound states. Moreover, we have also given the bound-state mass.
Next, we shall compare the above masses upon the symmetric vacuum with the ones
in the literature.

\section{Schwinger Boson and Its Bound State}
\label{4}

As pointed out in the introduction, the (1+1)-D MsGFT Eq.(1) with $\beta^2=4\pi
$ is equivalent to the massive Schwinger model at zero charge sector. The
Lagrangian of the latter is \cite{1,6,9,11,12} \cite{27} (A. Carroll $et \ al$)
\begin{equation}
{\cal L}=-{\frac {1}{4}}F_{\mu\nu}F^{\mu\nu}
        + {\bar \psi}_x[\gamma^{\mu}(i\partial_\mu - eA_{\mu})-m_f]\psi_x
\end{equation}
with $F_{\mu\nu}=\partial_{\mu}A_{\nu}-\partial_{\nu}A_{\mu}$. The
normal-ordering Hamiltonian corresponding to Eq.(23) with normal-ordering mass
$m_f$ is equal to Eq.(2) with the normal-ordering mass $m_0$. The
correspondence between the parameters in Eqs.(1) and (23) is
\begin{equation}
m_0^2={\frac {e^2}{\pi}}, m^2=2e^{\gamma}m_0 m_f \;,
\end{equation}
where $\gamma$ is the Euler constant. Eq.(23) is (1+1)-D QED with massive
fermion, and is obtained by inserting an additional mass term $m_f{\bar\psi}
\psi$ in the Lagrangian of the Schwinger model \cite{37}. The Schwinger model
was exactly solved \cite{37}, and shares some nontrivial properties with QCD
such as nontrivial vacuum, quark trapping, and so on. It is equivalent to a
massive free-scalar-field theory with the boson mass $m_0$ . For the particle
spectrum of this model, there exist no single-femion states but free boson
which is femion-antifemion bound state ( usually called Schwinger boson ).
Having an additional mass term, the massive Schwinger model has Schwinger boson
and its various excited states. In view of those aspects of the Schwinger
model, most investigations of the massive Schwinger model were involved in
tackling the confinement and particle spectrum of it, and mainly in the
small-$m_f$ effects. To our knowledge, except for a lattice study ( not
including those on light cone ), only Refs.~\cite{26,28} gave the masses of the
Schwinger boson and its bound state for a finite $m_f$. In this section, we
shall give the masses of the Schwinger boson and two-Schwinger-boson bound
state from the symmetric-vacuum results in the last section, and compare them
with the recent results in Refs.~\cite{29,30}. As for the vacuum structure, if
$m_f$ in Eq.(23) is greater than zero or $m_f$ is small, there exists no
asymmetric vacua according to the results in section $II$ ( $m_f>0$ implies
$m^2>0$, and hence $\tilde{m}_0^2<1$ from Eq.(8) ).

We first consider the Schwinger boson. From Eq.(19), one can have the Schwinger
boson mass $m_s$ upon a symmetric vacuum ($\varphi_0=0$)( taking $M=m_0$ )
\begin{equation}
\tilde{m}_s=e^\gamma \tilde{m}_f+\sqrt{e^{2\gamma}{\tilde{m}_f}^2+1}
\end{equation}
with $\tilde{m}_s={\frac {m_s}{m_0}}$ and $\tilde{m}_f={\frac {m_f}{m_0}}$.
When $\tilde{m}_f$ is infinitesimal, we obtain from the last equation
\begin{equation}
{\tilde{m}_s}^2=1+2e^\gamma \tilde{m}_f+2e^{2\gamma}{\tilde{m}_f}^2
           +O(\tilde{m}_f^3) \;.
\end{equation}
Performing fermion-mass perturbation technique for the massive Schwinger model
in the ``near'' light front coordinate system, Ref.~\cite{29} gave the
Schwinger boson mass to second order of $m_f$. We find that the first two terms
in the $r.h.s$ of the last equation is identical to the corresponding terms in
Eq.(3.16) of Ref.~\cite{29}, and the only difference is that the constant
factor in the $m_f^2$ term is $2e^{2\gamma}$ for our result but $e^{2\gamma}$
for Eq.(3.16) in Ref.~\cite{29}. More recently, Ref.~\cite{30} also gave almost
the identical result of the Schwinger boson mass up to second order of $m_f$
with that in Ref.~\cite{29}. Thus, for an infinitesimal $m_f$, our result of
the Schwinger boson mass has a good agreement with the ones in
Refs.~\cite{29,30}. For an illustration, a plot of our result and the results
in Ref.~\cite{29,30} are shown in Fig.7. In this figure, the results in
Refs.~\cite{29,30} are represented by dashed curves and coincide with each
other. This figure indicates that with the increase of $\tilde{m}_f$, our
result (solid curve) is more and more higher than the dashed curve, while for
small $\tilde{m}_f<0.2$, the solid curve nearly coincides with the dashed
curve.

Now we are in a position to discuss the two-Schwinger-boson bound state. From
Eq.(22), we gain the mass of the two-Schwinger-boson state upon the symmetric
vacuum $m_{sb}$
\begin{equation}
\tilde{m}_{sb}={\frac {{\tilde{m}_f}^2}{2\tilde{m}_s}}
[{\frac {1}{\sqrt{1-\tilde{m}_{sb}^2}}}
tan^{-1}\sqrt{{\frac {1+\tilde{m}_{sb}^2}{1-\tilde{m}_{sb}^2}}}
-{\frac {\pi}{4}}]
\end{equation}
with the reduced mass $\tilde{m}_{sb}\equiv {\frac {m_{sb}}{2m_s}}$.
From  Eqs.(26) and (27), we depict $\tilde{m}_{bs}$ with respect to the small
reduced fermion-mass $\tilde{m}_f$ in Fig.8 (the solide curve). In this figure,
the dashed curve is the corresponding result in Ref.~\cite{30} to second order
of the fermion mass $m_f$. This figure demonstrates that for a small
$\tilde{m}_f$ the result from the GWFA agree very well with the second-order
result of fermion-mass perturbation in Ref.~\cite{30}.

According to Refs.~\cite{26,27,28,29,30}, the mass-perturbation results have a
good agreement with analytical or numerical ones from many other techniques.
Therefore, we can say that for a small mass $m_f$, the GWFA give the masses of
Schwinger boson and two Schwinger-boson bound state at a good accuracy.

By the way, for finite fermion mass, the mass $\tilde{m}_s$ from Eq.(25) is
nearly linear in terms of $\tilde{m}_f$. We notice that in Ref.~\cite{26}, the
result about the mass of the Schwinger boson is also linear one in terms of the
$m_f$. For any given value of $m_f$, there exist infinitely many results from
Ref.~\cite{26}. For example, for $\tilde{m}_f=2.0$, besides the results in
Figs.2 and 3, Ref.~\cite{26} gave the other four values of the Schwinger-boson
mass : 4.78,5.97,6.9 and 7.70. For this case $\tilde{m}_f=2.0$, Eq.(25) give
$\tilde{m}_s=7.262$, which is between the last two values 6.9 and 7.7. The
relative error is 6 percent or so.

\section{Conclusion}
\label{5}

In this paper, we investigated the MsGFT with the GWFA in $(1+1)$ dimensions.
We discussed the ground, one- and two-particle states. For the ground
state, we demonstrate the existence of the asymmetric vacuum, obtain the
constraint of the coupling $\beta^2<8\pi$, and give the parameter regions of
the symmetric and the asymmetric vacua ( Fig.1 ). For the one-particle state,
the implicit formula Eq.(19) is obtained for the mass of a single MsG-particle.
As for the two-particle state, we show that the two-particle bound state can
exist upon an asymmetric vacuum. We also give the bound-state mass formula
Eq.(22) as well as the model-parameter regions of the bound stats upon the
symmetric and the asymmetric vacua, and discuss the dependence of the
bound-state masses upon the model parameters $\beta^2$ and $\tilde{m}_0^2$.
Finally, the masses of the Schwinger boson and the two-Schwinger-boson bound
state in the massive Schwinger model are calculated according to Eqs.(19) and
(22), and have a good agreement with those in the recent literature
\cite{29,30}( Fig.7 and 8 ) when the fermion mass $m_f$ is small. 

  Before closing the paper, we want to give a further discussion about Fig.1
and Fig.3. Fig.1 is just the phase diagram of the vacuum. In this figure, with
continuous variations in $\beta^2$ and $\tilde{m}_0^2$, a symmetric phase of
the vacuum can transit the long-dashed boundary to an asymmetric phase. This
implies the occurence of a phase transition. Furthermore, Fig.3 is an explicit
illustration of the GEP with $\tilde{m}_0^2=20$, and it indicates that the
vacuum average value of the field operator $\varphi_0$ changes discontinously
from zero to nonzero when $\beta^2$ increases gradually. That is to say, the
GWFA predicts a first-order phase transition in the MsGFT. Nevertheless, it
could be inadvisable to conclude that a true first-order phase transition
occurs in the MsGFT. As mentioned in section II, some GWFA information related
to the phase transition may be changed by a better approximation method. The
GWFA is a simple non-perturbative approach. Although it is effective and useful
for investigating many problems or phenomena, we should not to expect too much
of it, particularly when we are concerned with a phase transition. In fact, for
a few (1+1)-D field theories, the GWFA predicts the wrong order of the phase
transition. We take the $\lambda\phi^4$ field theory as the first example. B.
Simon and R. B. Griffiths gave a rigorous theorem that for the (1+1)-D
$\lambda\phi^4$ field theory in the presence of an external field $B\ne 0$, the
vacuum of it is unique \cite{38}. Further, S. J. Chang proved that the
occurence of the first order phase transition in the (1+1)-D $\lambda\phi^4$
field theory violates Simon-Griffiths theorem, but a second order phase
transition can be compatible with this theorem \cite{23}. As is known, the GWFA
predicts just a first order phase transition in this theory \cite{23,21}, and a
second order phase transition can occur in the (1+1)-D $\lambda\phi^4$ field
theory \cite{23,39}. That is to say, the GWFA predicts correctly the existence
of the phase transition in the (1+1)-D $\lambda\phi^4$ field theory, and
predicts incorrectly just the nature of the phase transition. Another example
is the (1+1)-D $\phi^6$ field theory. The GWFA predicts a first order phase
transition in this theory \cite{16}. But a coupled-cluster-method investigation
indicated that for the region where the two-particle bound state exists, a
first order phase transition can occur in the (1+1)-D $\phi^6$ field theory,
nevertheless, the critical curve is different from the corresponding one in the
GWFA result \cite{40}. Additionally, for the region where a two-particle bound
state disappears, no first order phase transitions exist, but a second order
phase transition is believed to occur in the (1+1)-D $\phi^6$ field theory
\cite{40}. In view of these situations of the above two theories, we feel that
Simon-Griffiths theorem could perhaps have an effect on the other (1+1)-D field
theories to some extent. Therefore, we conjecture that for the (1+1)-D MsGFT,
the GWFA predict correctly the existence of a phase transition, but could make
a mistake in predicting the critical boundary and the nature of the transition,
perhaps which be similar to those in the (1+1)-D $\phi^6$ field theory. In
order to get a definite conclusion, some better approximate methods should be
used \cite{22,39,40}. We believe that after considering the higher order
correction of the GEP \cite{22,39,40}, one maybe obtain different figures from
Fig.1 and Fig.3, but the asymmetric vacuum would still exist. In a general,
when a classical vacuum in (1+1) dimensions is spontaneously symmetry-broken
for some region of the model parameters, quantum effects are not suficient to
turn the vacuum completely symmmetrical for all values of the model parameters.
The existence of the asymmetric vacuum should be reasonable.

  Besides, in Fig.1, when $\tilde{m}_0^2$ tends to zero, the symmetric phase of
the vacuum becomes a dynamical-symmetry-breakdown phase. This perhaps implies
the existence of a phase transition. In fact, Ref.~\cite{1} ( the end paragraph
on page 407 in the book ) pointed out that there may be a phase transition (
and long range order ) if $m_0$ is small enough. Additionally, as mentioned in
section II, the constraint of $\beta^2<8\pi$ in the MsGFT is the same as in the
sGFT. It is well known that $\beta^2=8\pi$ is a critical point at which the
Kosterlitz-Thouless transition occurs in the sG system \cite{41}. However, when
$\tilde{m}_0^2\ne 0$, $\beta^2=8\pi$ could not suggests the Kosterlitz-Thouless
transition and the zero mass phase could not exist, for $\mu=0$ can not give
either the local or the global minimum of the energy densty but the infinity
(see section II). As a matter of fact, the disappearance of the massless
Kosterlitz-Thouless phase has been shown in Ref.~\cite{8}( 1994 ). By the
way, if one use the GWFA with a finite momentum cutoff \cite{42}\cite{18}
( Phys. Rev. B ), it is possible to have a satisfactory understanding about the
problems discussed in this paragraph.

  In conclusion, the results in this paper are qualitatively correct and are
necessary and helpful for further investigations of the MsGFT at least. The
GWFA results about the ground state of the MsGFT is useful for the Yukawa gas
and the lattice Abelian Higgs model, and meanwhile discussions about other
properties or phenomena of the massive Schwinger model with the GWFA will be
also interesting and useful.

\acknowledgments
The author would like to acknowledge professor Guang-jiong Ni and professor
Bo-wei Xu for their helpful discussions. A part of this work was done while
the author worked at Fudan University. This project was supported jointly by
the Municipal Postdoctoral Foundation of Shanghai, the National Postdoctoral
Foundation of China, the President foundation of Shanghai Jiao Tong University,
and the National Natural Science foundation of China with grant No. 19875034.

\figure{Fig.1  The $\beta^2$---$\tilde{m}_0^2$ parameter space for
          the massive sine-Gordon field theory in (1+1) dimensions, which is
          plotted from Eqs.(10)$-$(13). The vacuum is symmetric in the domain
          $I$, and is asymmetric both in the domain $II$ without bound state
          and in the domain $III$ with bound state. The short-dashed curve
          corresponds to $cos(\beta\varphi_0)=0$ and the dot-dashed curve to
          the vacuum at $\varphi_0=0.75$. }
\figure{Fig.2  The reduced GEP of the massive sine-Gordon field theory in (1+1)
           dimensions for the case of $\tilde{m}_0^2=1.5$. Only one-half of
           the symmetric potential is shown. In this figure, curves from the
           left to the right are drawn for $\beta^2=0.0004,\; 0.25,\; 1$
           and $25$, respectively.}
\figure{Fig.3  Similar to Fig.2 but in $\tilde{m}_0^2=20$.
        In this figure, curves from the highest to the lowest are drawn
        for $\beta^2=1.44, \;2.25, \; \beta_c\approx 3.03, \; 4.0$
         and $4.84$,respectively.}
\figure{Fig.4  The dependence of the reduced mass of the bound state in the
       symmetric vacuum upon $\beta^2$ for some values of $\tilde{m}_0^2$. In
       this figure, Curves from the lowest to the highest are drawn for
       $\tilde{m}^2_0=0.05,\; 0.2,\; 0.4,\; 0.6,\; $ and $0.8$, respectively.}
\figure{Fig.5  The dependence of the reduced mass of the bound state in the
        asymmetric vacuum $\varphi_0=0.75$ upon $\beta^2$.}
\figure{Fig.6 The dependence of the reduced mass of the bound state in the
       asymmetric vacuum $\varphi_0=0.75$ upon $\tilde{m}_0^2$.}
\figure{Fig.7  The comparison of the Schwinger boson mass Eq.(26) (solid curve)
       with the corresponding second-order results in Refs.~\cite{29,30}(dashed
       curve, coincided).}
\figure{Fig.8 The dependence of the reduced mass of the two-Schwinger-boson
        bound state in the symmetric vacuum $\tilde{m}_{bs}$ upon the reduced
        mass $\tilde{m}_f$. The dashed curve is the corresponding result in
        Ref.~\cite{30}.}

\end{document}